%% file: SsdTaylor.tex
\documentclass[preprint,12pt]{elsarticle}
\usepackage{graphics,epsfig}
\usepackage{graphicx}
\usepackage{amssymb,amstext,amsmath}

\usepackage{booktabs}
\usepackage{natbib}
\usepackage[latin1]{inputenc}
\usepackage{epstopdf}
\begin{document}
\input{def.tex}
\begin{frontmatter}
\title{Continuum dislocation theory accounting for statistically stored dislocations and Taylor hardening}
\author{K. C. Le\footnote{corresponding author: ++49 234 32-26033, email: chau.le@rub.de}$^a$, P. Sembiring$^a$, and T. N. Tran$^b$}
\address{$^a$ Lehrstuhl f\"{u}r Mechanik - Materialtheorie, Ruhr-Universit\"{a}t Bochum, D-44780 Bochum, Germany
\\
$^b$ Computational Engineering, Vietnamese-German University, Binh Duong, Viet nam}
\begin{abstract}	
This paper develops the small strain continuum dislocation theory accounting for statistically stored dislocations and Taylor hardening for single crystals. As illustration, the problem of anti-plane constrained shear of single crystal deforming in single slip is solved within the proposed theory. The distribution of geometrically necessary dislocations in the final state of equilibrium as well as the stress-strain curve exhibiting the Bauschinger translational work hardening and the size effect are found. Comparison with the stress-strain curve obtained from the continuum dislocation theory without statistically stored dislocations and Taylor hardening is provided.
\end{abstract}

\end{frontmatter}

\section{Introduction}
\label{intro}

Macroscopically observable plastic deformations in single crystals and polycrystalline materials are caused by nucleation, multiplication and motion of geometrically necessary dislocations (GNDs). There are various reasonable experimental evidences supporting the so-called low energy dislocation structure (LEDS) hypothesis formulated first by \citet{Hansen1986}: dislocations appear in the crystal lattice to reduce its energy (see also  \citep{Laird1986,Kuhlmann1989}). Motion of dislocations yields the dissipation of energy which, in turn, results in a resistance to the dislocation motion. The general structure of continuum dislocation theory (CDT) must therefore reflect this physical reality: energy decrease by nucleation of GNDs and resistance to the motion of GNDs due to dissipation. Just in recent years various phenomenological models of crystals with continuously distributed dislocations which are able to predict the density of GNDs as well as the accompanying size effects have been proposed in \citep{Acharya2001,Gurtin2002,Gurtin2007,Berdichevsky2006a,Berdichevsky2006b,Berdichevsky2007,Le2008a,Le2008b,Le2009,Kochmann2008,Kochmann2009a,Kochmann2009b,Le2010a,Kaluza2011,Le2012,Le2013} (see also the finite strain CDT proposed by \citet{Le1996a,Le1996b,Le1996c,Ortiz1999,Ortiz2000,Le2014,Koster2015}).

In addition to the geometrically necessary dislocations there exists another family of dislocations which does not show up in the macroscopically observable plastic slip but nevertheless may have significant influences on the nucleation of GNDs and the work hardening of crystals. For any closed circuit surrounding an area, which is regarded as infinitesimal compared with the characteristic size of the macroscopic body but may still contains a large number of dislocations, the resultant Burgers vector of these dislocations always vanishes, so the closure failure caused by the incompatible plastic slip is not affected by them. Following \citet{Ashby1970} we call these dislocations statistically stored dislocations (SSDs). As a rule, the statistically stored dislocations in unloaded crystals at low temperatures exist in form of dislocation dipoles in two-dimensional case or small planar dislocation loops whose size is comparable with the atomic distance in three-dimensional case. The simple reason for this is that the energy of a dislocation dipole (or a small planar dislocation loop) is much smaller than that of dislocations apart, so the bounded state of dislocations renders low energy to the whole crystal. From the other side, due to their low energy, the dislocation dipoles (loops) can easily be created (as well as annihilated) by thermal fluctuations. The statistically stored dislocations play two important roles in the plastic deformations of crystals: i) together with the Frank-Read source \citep{Hirth1968} they provide additional sources for the nucleation of GNDs due to the fact that, when the applied shear stress becomes large enough, the dislocation dipoles dissolve to form the freely moving GNDs, ii) the neutral dipoles (loops) of SSDs act as obstacles that impede the motion of GNDs leading to the nonlinear work hardening. In view of their important roles in ductile crystals, the account of SSDs in the CDT would make the material models more realistic. \citet{Arsenlis2004} proposed a set of evolution equations for the densities of GNDs and SSDs within the crystal plasticity. The density of SSDs evolves through Burgers vector-conserving reactions, while that of GNDs evolves due to the divergence of dislocation fluxes. Except the missing thermal fluctuations in the nucleation of SSDs, it was also unclear whether such an approach could be related to the energetics of crystals containing dislocations and the LEDS-hypothesis mentioned above. \citet{Berdichevsky2006a} was the first who included the density of SSDs in the free energy density of the crystal. However, to the best of our knowledge, the more pronounced influence of the SSDs on the yield stress and the dissipation within the CDT has not been considered up to now. This paper aims at filling this gap. Its main idea is to propose the dissipation function depending on both densities of GNDs and SSDs in such a way that the obtained yield stress combines the constant plastic yield stress due to the Peierls barrier and the Taylor contribution that is proportional to the square root of the total density of GNDs and SSDs \citep{Taylor1934}. Then we apply the proposed theory to the problem of anti-plane constrained shear. We solve this problem numerically and find the distribution of GNDs in the final state of equilibrium as well as the stress-strain curve. We show the size effect for the threshold stress, the nonlinear work hardening due to the combined GNDs and SSDs, and the Bauschinger effect for the loading, elastic unloading, and loading in the opposite direction.

The paper is organized as follows. In Section 2 the kinematics of CDT taking into account GNDs and SSDs is laid down. Section 3 proposes the thermodynamic framework for the CDT with SSDs and Taylor hardening. In Section 4 the problem of anti-plane constrained shear is analyzed. Section 5 presents the numerical solution of this problem and discusses the distribution of GNDs, the stress-strain curve, the Bauschinger translational work hardening and the size effect. Finally, Section 6 concludes the paper.  

\section{Kinematics}
\label{sec:1}
In this paper we restrict ourselves to the small strain (or geometrically linear) continuum dislocation theory for single crystals. For simplicity we shall use some fixed rectangular cartesian coordinates and denote by $\mathbf{x}$ the position vector of a generic material point of the crystal. Kinematic quantities characterizing the observable deformation of this single crystal are the displacement field $\mathbf{u}(\mathbf{x})$ and the plastic distortion field $\bbeta (\mathbf{x})$ that is incompatible. For single crystals having $n$ active slip systems, the plastic distortion is in general given by
\begin{equation}
\label{eq:1.1}
\bbeta (\mathbf{x})=\sum_{\mathfrak{a}=1}^n \beta ^\mathfrak{a}(\mathbf{x}) \mathbf{s}^\mathfrak{a}\otimes \mathbf{m}^\mathfrak{a} \quad (\beta _{ij}=\sum_{\mathfrak{a}=1}^n \beta ^\mathfrak{a}s_i^\mathfrak{a}m_j^\mathfrak{a}),
\end{equation}  
with $\beta ^\mathfrak{a}$ being the plastic slip, where the pair of constant and mutually orthogonal unit vectors $\mathbf{s}^\mathfrak{a}$ and $\mathbf{m}^\mathfrak{a}$ is used to denote the slip direction and the normal to the slip planes of the corresponding $\mathfrak{a}$-th slip system, respectively. Thus, there are altogether $3+n$ degrees of freedom at each point of this generalized continuum. Here and later, equivalent formulas for the components of tensors are also given on the same line in brackets, where the Latin lower indices running from 1 to 3 indicate the projections onto the corresponding coordinates, while the Gothic upper index $\mathfrak{a}$ running from 1 to $n$ numerates the slip systems. We use Einstein's summation convention, according to which summation over a repeated Latin index from 1 to 3 is understood. We see immediately from \eqref{eq:1.1} that $\text{tr}\bbeta =\beta _{ii}=0$, so the plastic distortion is volume preserving. 

\begin{figure}[htb]
	\centering
	\includegraphics[height=7cm]{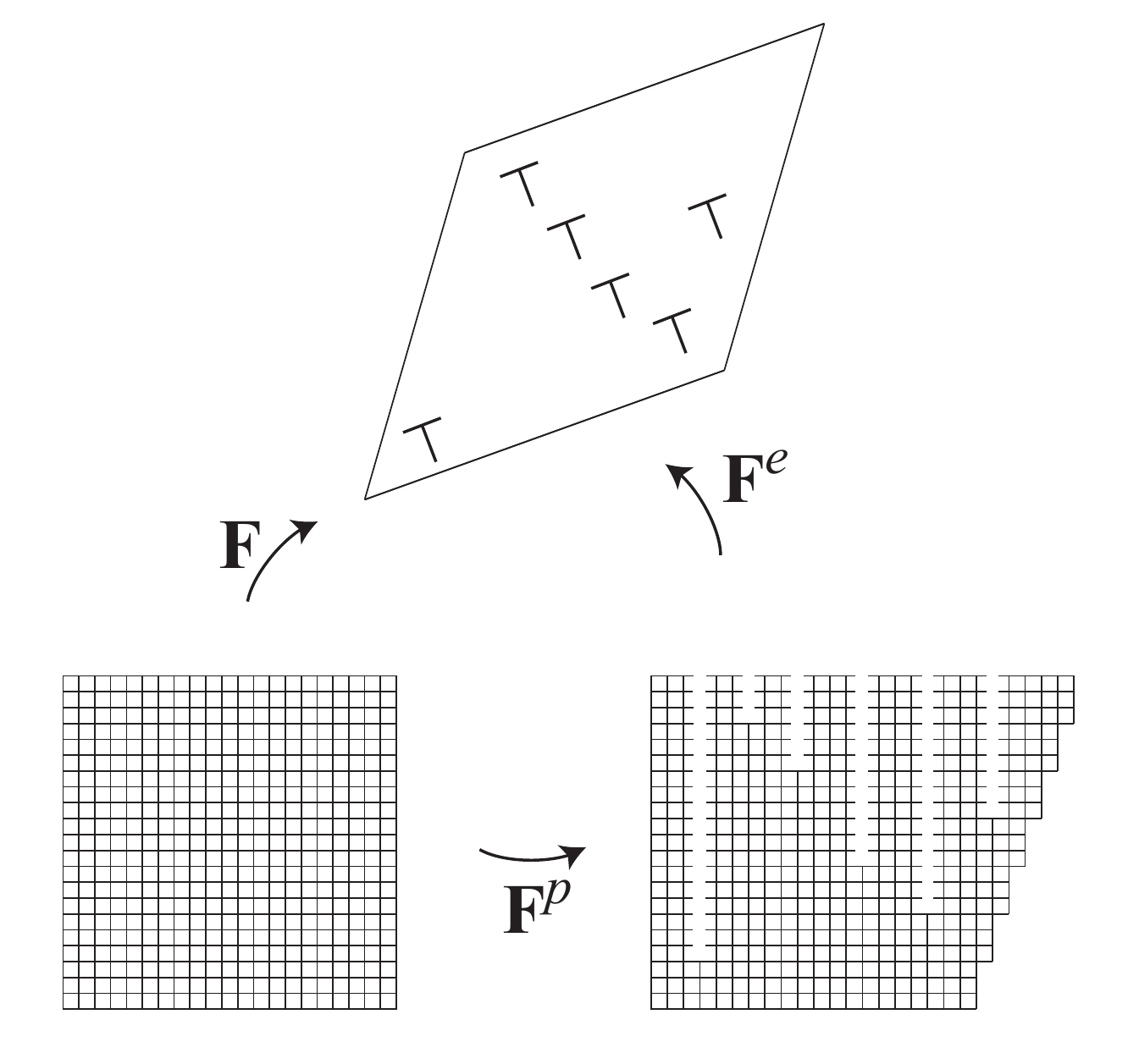}
	\caption{Additive decomposition}
	\label{fig:Nhslip}
\end{figure}

The total compatible strain tensor field can be obtained from the displacement field according to
\begin{equation}
\label{eq:1.2}
\bvarepsilon =\frac{1}{2}(\mathbf{u}\nabla +\nabla \mathbf{u}) \quad (\varepsilon _{ij}=\frac{1}{2}(u_{i,j}+u_{j,i})).
\end{equation}
The incompatible plastic strain tensor field is the symmetric part of the plastic distortion field
\begin{equation}
\label{eq:1.3}
\bvarepsilon ^p =\frac{1}{2}(\bbeta +\bbeta ^T) \quad (\varepsilon ^p_{ij}=\frac{1}{2}(\beta_{ij}+\beta_{ji})).
\end{equation}
For small strains we shall use the additive decomposition of the total strain into the elastic and plastic parts. Therefore, the elastic strain tensor field is equal to
\begin{equation}
\label{eq:1.4}
\bvarepsilon ^e=\bvarepsilon -\bvarepsilon ^p \quad (\varepsilon ^e_{ij}=\varepsilon _{ij}-\varepsilon ^p_{ij}).
\end{equation}
Although not absolutely necessary, we introduce for the illustration purpose the elastic distortion tensor field according to
\begin{equation*}
\bbeta ^e=\mathbf{u}\nabla -\bbeta \quad (\beta ^e_{ij}=u_{i,j}-\beta _{ij}).
\end{equation*}
The relationship between these three distortion fields is illustrated in Fig.~\ref{fig:Nhslip}, where $\mathbf{F}=\mathbf{I}+\mathbf{u}\nabla $, $\mathbf{F}^p=\mathbf{I}+\bbeta $, $\mathbf{F}^e=\mathbf{I}+\bbeta ^e$. Looking at this Figure we see that the plastic distortion $\bbeta $ is the distortion {\it creating} dislocations (either inside or at the boundary of the volume element) or {\it changing} their positions in the crystal without deforming the crystal lattice. In contrary, the elastic distortion $\bbeta^e$ deforms the crystal lattice having {\it frozen} dislocations \citep{Le2014}.

Since the lattice rotation can directly be measured by the Electron Back\-scatter Diffraction (EBSD) technique \citep{Kysar2010}, it is interesting to express it in terms of the displacement gradient and plastic slip. Using the additive decomposition for the displacement gradients $\mathbf{u}\nabla =\boldsymbol{\varepsilon }+\boldsymbol{\omega }$, we find the components of the total rotation tensor in the form
\begin{equation*}
\omega _{ij}=\frac{1}{2}(u_{i,j} -u_{j,i}).
\end{equation*}
This tensor is obviously skew-symmetric. The plastic rotation tensor is the skew-symmetric part of the plastic distortion
\begin{equation*}
\omega ^p_{ij}=\frac{1}{2}(\beta _{ij} -\beta _{ji})=\sum_{\mathfrak{a}=1}^n\frac{1}{2}\beta ^\mathfrak{a}(s^\mathfrak{a}_im^\mathfrak{a}_j-s^\mathfrak{a}_jm^\mathfrak{a}_i).
\end{equation*}
The elastic (lattice) rotation tensor is the difference between these two tensors
\begin{equation*}
\omega ^e_{ij}=\omega _{ij}-\omega ^p_{ij}=\frac{1}{2}(u_{i,j} -u_{j,i})-\sum_{\mathfrak{a}=1}^n\frac{1}{2}\beta ^\mathfrak{a} (s^\mathfrak{a}_im^\mathfrak{a}_j-s^\mathfrak{a}_jm^\mathfrak{a}_i).
\end{equation*}

To characterize the geometrically necessary dislocations (GNDs) belonging to one slip system let us consider one term $\bbeta ^\mathfrak{a}=\beta ^\mathfrak{a}\mathbf{s}^\mathfrak{a}\otimes \mathbf{m}^\mathfrak{a}$ in the sum \eqref{eq:1.1}. The measure of incompatibility of this plastic distortion, introduced by \citet{Nye1953}, \citet{Bilby1955}, and \citet{Kroener1955}, reads
\begin{equation*}
\balpha ^\mathfrak{a}=-\bbeta ^\mathfrak{a}\times \nabla \quad (\alpha _{ij}^\mathfrak{a}=\epsilon _{jkl}\beta ^\mathfrak{a}_{il,k}),
\end{equation*}
where $\epsilon _{ijk}$ is the permutation symbol. With $\bbeta ^\mathfrak{a}=\beta ^\mathfrak{a}\mathbf{s}^\mathfrak{a}\otimes \mathbf{m}^\mathfrak{a}$ we have
\begin{equation}
\label{eq:1.6}
\balpha ^\mathfrak{a}= \mathbf{s}^\mathfrak{a}\otimes (\nabla \beta ^\mathfrak{a} \times \mathbf{m}^\mathfrak{a}) \quad (\alpha ^\mathfrak{a}_{ij}= s_i^\mathfrak{a} \epsilon _{jkl} \beta ^\mathfrak{a}_{,k}m_l^\mathfrak{a}).
\end{equation}
To find the number of dislocations per unit area of the $\mathfrak{a}$-th slip system we take an infinitesimal area $da$ with the unit normal vector $\mathbf{n}$. Then the resultant Burgers vector of those geometrically necessary dislocations belonging to this slip system, whose dislocation lines cross the area $da$ is given by
\begin{equation*}
\mathbf{b}^\mathfrak{a}=\balpha ^\mathfrak{a} \cdot \mathbf{n}\, da \quad (b^\mathfrak{a}_i=\alpha ^\mathfrak{a}_{ij}n_j\, da).
\end{equation*}
This is quite similar to the Cauchy formula relating the traction with the stress tensor. With $\balpha ^\mathfrak{a}=\mathbf{s}^\mathfrak{a}\otimes (\nabla \beta ^\mathfrak{a} \times \mathbf{m}^\mathfrak{a})$ we get
\begin{equation*}
\mathbf{b}^\mathfrak{a}=\mathbf{s}^\mathfrak{a} ((\nabla \beta ^\mathfrak{a} \times \mathbf{m}^\mathfrak{a})\cdot \mathbf{n})\, da = \mathbf{s}^\mathfrak{a} (\nabla \beta ^\mathfrak{a} \cdot (\mathbf{m}^\mathfrak{a}\times \mathbf{n}))\, da .
\end{equation*}
Provided the direction $\mathbf{t}^\mathfrak{a}$ tangential to the dislocation lines lying in the slip planes is known for the $\mathfrak{a}$-th slip system, then one can choose the infinitesimal area $da$ with the unit normal vector $\mathbf{t}^\mathfrak{a}$ to compute the resultant Burgers vector of those geometrically necessary dislocations, whose dislocation lines cross this area under the right angle
\begin{equation*}
\mathbf{b}^\mathfrak{a}=\mathbf{s}^\mathfrak{a} (\nabla \beta ^\mathfrak{a} \cdot (\mathbf{m}^\mathfrak{a}\times \mathbf{t}^\mathfrak{a}))\, da =\mathbf{s}^\mathfrak{a} \partial_\nu \beta ^\mathfrak{a}\, da,
\end{equation*}
where $\partial _\nu \beta ^\mathfrak{a}=\nabla \beta ^\mathfrak{a} \cdot \bnu ^\mathfrak{a}$, with $\bnu ^\mathfrak{a}=\mathbf{m}^\mathfrak{a}\times \mathbf{t}^\mathfrak{a}$ being the vector lying in the slip plane and perpendicular to $\mathbf{t}^\mathfrak{a}$. This resultant Burgers vector can be decomposed into the sum of two Burgers vectors: one parallel to $\mathbf{t}^\mathfrak{a}$ representing the screw dislocations, and another perpendicular to $\mathbf{t}^\mathfrak{a}$ corresponding to the edge dislocations. Therefore, the number of dislocations per unit area of each sort is given by
\begin{equation}
\label{eq:1.7}
\rho _{\odot}^\mathfrak{a}=\frac{1}{b}|\mathbf{s}^\mathfrak{a}\cdot \mathbf{t}^\mathfrak{a}||\partial_\nu \beta ^\mathfrak{a}|, \quad \rho _{\|}^\mathfrak{a}=\frac{1}{b}|\mathbf{s}^\mathfrak{a}\cdot \bnu^\mathfrak{a}||\partial_\nu \beta ^\mathfrak{a}| ,
\end{equation}
where the symbol $\odot $ indicates the screw dislocations, while $\|$ the edge dislocations.

In addition to the geometrically necessary dislocations there exist another family of dislocations which does not show up in the macroscopically observable plastic distortion but nevertheless may have significant influences on the nucleation of GNDs and the work hardening of crystals. For any closed circuit surrounding an infinitesimal area (in the sense of continuum mechanics) the resultant Burgers vector of these dislocations always vanishes, so the closure failure caused by the incompatible plastic distortion is not affected by them. Following \citet{Ashby1970} we call these dislocations statistically stored dislocations (SSDs). As a rule, the statistically stored dislocations in unloaded crystals at low temperatures exist in form of dislocation dipoles (in two-dimensional case) or small planar dislocation loops, whose size is comparable with the atomic distance (in three-dimensional case) (\citet{Arsenlis1999} have found also other three-dimensional self-terminating dislocation structures with zero net Burgers vector). The simple reason for this is that the energy of a dislocation dipole (or a small dislocation loops) is much smaller than that of dislocations apart, so this bounded state of dislocations renders low energy to the whole crystal. From the other side, due to their low energy, the dislocation dipoles can easily be created (as well as annihilated) by thermal fluctuations. As can be shown by the statistical mechanics of dislocations in two-dimensional case \citep{Berdichevsky2002}, the number of such dipoles remains nearly constant at the constant temperature. Let us denote the density of SSDs for each slip system by $\rho ^\mathfrak{a}_{st}$. 

\section{Thermodynamic framework}
\label{sec:2}

To set up phenomenological models of crystals with continuously distributed dislocation using the methods of non-equilibrium thermodynamics for irreversible processes let us begin with the free energy density. As a function of the state, the free energy density may depend only on the state variables. Following \citet{Kroener1992} we will assume that the elastic strain $\bvarepsilon^e$, the densities of GNDs $\balpha ^\mathfrak{a}$ ($\mathfrak{a}=1,\ldots ,n$), and the absolute temperature $T$ characterize the current state of the crystal, so these quantities are the state variables of the continuum dislocation theory. The reason why the plastic distortion $\bbeta $ cannot be qualified for the state variable is that it depends on the cut surfaces and consequently on the whole  history of creating dislocations (for instance, climb or glide dislocations are created quite differently). Likewise, the gradient of plastic strain tensor $\bvarepsilon^p$ cannot be used as the state variable by the same reason. In contrary, the dislocation densities $\balpha ^\mathfrak{a}$ depend only on the characteristics of GNDs in the current state (Burgers vector and positions of dislocation lines) and not on how they are created, so $\balpha ^\mathfrak{a}$ is the proper state variable. In addition to these state variable one should include also the densities of statistically stored dislocations $\rho ^\mathfrak{a}_{st}$ into the list of state variables. However, at low temperature these statistically stored dislocations prefer to exist in form of dislocation dipoles to render the crystal a low energy. Provided the density of such dipoles depends only on the temperature, their energy contribution is a constant that can be omitted. Besides, due to the charge neutrality of dipoles, the energy of interaction between SSDs and GNDs are negligibly small compared to the energy contributions of GNDs. Thus, if we consider isothermal processes of deformation, then the free energy per unit volume of crystal (assumed as macroscopically homogeneous) must be a function of $\bvarepsilon^e$ and $\balpha ^\mathfrak{a}$ 
\begin{equation}\label{eq:2.1}
\psi = \psi (\bvarepsilon^e,\balpha ^\mathfrak{a}).
\end{equation}

With this free energy density we can now write down the energy functional of the crystal. Let the undeformed single crystal occupy the region $\mathcal{V}$ of the three-dimensional euclidean point space. The boundary of this region, $\partial \mathcal{V}$, is assumed to be the closure of union of two non-intersecting surfaces, $\partial _k$ and $\partial _s$. Let the displacement vector $\mathbf{u}(\mathbf{x})$ be a given smooth function of coordinates, and, consequently, the plastic slips $\beta ^\mathfrak{a}(\mathbf{x})$ vanish 
\begin{equation}\label{eq:2.2}
\mathbf{u}(\mathbf{x})=\tilde{\mathbf{u}}(\mathbf{x}),\quad \beta ^\mathfrak{a}(\mathbf{x})=0 \quad \text{for $\mathbf{x}\in \partial _k$}.
\end{equation}
At the remaining part $\partial _s$ of the boundary the surface load (traction) $\mathbf{f}$ is specified. If no body force acts on this crystal, then its energy functional is defined as
\begin{equation}
\label{eq:2.3}
I[\mathbf{u}(\mathbf{x}),\beta ^\mathfrak{a}(\mathbf{x})]=\int_{\mathcal{V}}\psi (\bvarepsilon^e,\balpha ^\mathfrak{a})\, dx-\int_{\partial _s} \mathbf{f}\cdot \mathbf{u}\, da,
\end{equation}
with $dx=dx_1dx_2dx_3$ denoting the volume element and $da$ the area element. Provided the resistance to the dislocation motion can be neglected, then the following variational principle turns out to be valid for single crystals: the true displacement field $\check{\mathbf{u}}(\mathbf{x})$ and the true plastic slips $\check{\beta }^\mathfrak{a}(\mathbf{x})$ in the {\it final} state of deformation in equilibrium minimize energy functional \eqref{eq:2.3} among all admissible fields satisfying the constraints \eqref{eq:2.2}.

Let us find out the necessary conditions which must be satisfied by the energy minimizer in the final state of deformation in equilibrium. Taking the first variation of functional \eqref{eq:2.3} we have
\begin{equation*}
\delta I=\int_{\mathcal{V}}[\sigma _{ij}(\delta \varepsilon _{ij}-\delta \varepsilon ^p_{ij})+\sum_{\mathfrak{a}=1}^n \kappa ^\mathfrak{a} _{ij}\delta \alpha ^\mathfrak{a} _{ij}]\, dx
-\int_{\partial _s} f_i \delta u_i\, da,
\end{equation*}
where
\begin{equation}
\label{eq:2.4}
\sigma _{ij}=\frac{\partial \psi }{\partial \varepsilon ^e_{ij}},\quad \kappa ^\mathfrak{a} _{ij}=\frac{\partial \psi }{\partial \alpha ^\mathfrak{a}_{ij}}, \quad \mathfrak{a}=1,\ldots ,n.
\end{equation}
We call $\bsigma $ (Cauchy) stress tensor, while $\bkappa ^\mathfrak{a}$ higher order stress tensors. Taking the symmetry of $\bsigma $ into account and using the kinematic relations \eqref{eq:1.1}-\eqref{eq:1.6}, we obtain
\begin{equation*}
\delta I=\int_{\mathcal{V}}(\sigma _{ij}\delta u_{i,j}-\sigma _{ij}\sum_{\mathfrak{a}=1}^n \delta \beta ^\mathfrak{a} s_i^\mathfrak{a}m_j^\mathfrak{a}+\sum_{\mathfrak{a}=1}^n \kappa ^\mathfrak{a} _{ij}s_i^\mathfrak{a} \epsilon _{jkl} \delta \beta ^\mathfrak{a}_{,k}m_l^\mathfrak{a})\, dx
-\int_{\partial _s} f_i \delta u_i\, da.
\end{equation*}
Integration by parts with the use of the kinematic boundary conditions \eqref{eq:2.2} causing $\delta u_i=0$ and $\delta \beta ^\mathfrak{a}=0$ on $\partial _k$ yields
\begin{align*}
\delta I&=\int_{\mathcal{V}}[-\sigma _{ij,j}\delta u_{i}-\sum_{\mathfrak{a}=1}^n(\tau ^\mathfrak{a}+s_i^\mathfrak{a}\kappa ^\mathfrak{a} _{ij,k}\epsilon _{ljk}m_l^\mathfrak{a})\delta \beta ^\mathfrak{a}]\, dx
\\
&+\int_{\partial _s} [(\sigma _{ij}n_j-f_i) \delta u_i+\sum_{\mathfrak{a}=1}^n s_i^\mathfrak{a}\kappa ^\mathfrak{a} _{ij}\epsilon _{ljk}n_km_l^\mathfrak{a}\delta \beta ^\mathfrak{a}]\, da,
\end{align*}
where $\tau ^\mathfrak{a}=s_i^\mathfrak{a}\sigma _{ij}m_j^\mathfrak{a}$ is the resolved shear stress (Schmid stress). Since the variations $\delta u_{i}$ and $\delta \beta ^\mathfrak{a}$ can be chosen arbitrarily in $\mathcal{V}$ and on $\partial _s$, from $\delta I=0$ follow the equilibrium equations
\begin{equation}
\label{eq:2.5}
\sigma _{ij,j}=0, \quad s_i^\mathfrak{a}\kappa ^\mathfrak{a} _{ij,k}\epsilon _{ljk}m_l^\mathfrak{a}+\tau ^\mathfrak{a}=0, \quad \mathfrak{a}=1,\ldots ,n
\end{equation}
and the boundary conditions at $\partial _s$
\begin{equation}
\label{eq:2.6}
\sigma _{ij}n_j=f_i,\quad s_i^\mathfrak{a}\kappa ^\mathfrak{a} _{ij}\epsilon _{ljk}n_km_l^\mathfrak{a}=0, \quad \mathfrak{a}=1,\ldots ,n.
\end{equation}
The first equation of \eqref{eq:2.5} corresponds to the equilibrium of macro-forces acting on the volume element of crystal, while $n$ remaining equations express the equilibrium of micro-forces acting on dislocations of the corresponding slip system. The first terms in \eqref{eq:2.5}$_2$ are called the back-stresses which are nothing else but the resultant forces acting on a dislocation from other dislocations. Substituting \eqref{eq:2.4} into \eqref{eq:2.5}, we get the closed system of $3+n$ governing equations to determine $3+n$ unknown functions.

In real crystals there is however always the resistance to the dislocation motion causing the energy dissipation that changes the above variational principle as well as the equilibrium conditions. Various factors like impurities, inclusions, grain boundaries et cetera may have influence on this resistance to dislocation motion in polycrystals. However, in single crystals we may count two main sources. The first one is the periodic energy landscape of discrete crystal lattice (Peierls barriers) that, in combination with the small viscosity, leads to the rate independent dissipation \citep{Puglisi2005}. The second one is dislocations themselves acting as obstacles (dislocation forest) leading to the Taylor hardening \citep{Taylor1934}. So the dissipation potential can be proposed in the form
\begin{equation}
\label{eq:2.7}
D=D(\rho _{\odot}^\mathfrak{a},\rho _{\|}^\mathfrak{a},\rho ^\mathfrak{a}_{st},\dot{\beta }^\mathfrak{a}).
\end{equation}
We assume that this dissipation potential is the homogeneous function of the first order with respect to the plastic slip rates $\dot{\beta }^\mathfrak{a}$. The simplest version of this dissipation potential would be
\begin{equation}\label{eq:2.7a}
D=\sum_{\mathfrak{a}=1}^n g^\mathfrak{a}(\rho _{\odot}^\mathfrak{a},\rho _{\|}^\mathfrak{a},\rho ^\mathfrak{a}_{st}) |\dot{\beta }^\mathfrak{a}|,
\end{equation}
where $g^\mathfrak{a}(\rho _{\odot}^\mathfrak{a},\rho _{\|}^\mathfrak{a},\rho ^\mathfrak{a}_{st})$ are positive functions of the dislocation densities. In this case the latent hardening due to the cross-slip is neglected. The more complicated model taking the latent hardening into account is
\begin{equation*}
D=\sum_{\mathfrak{a},\mathfrak{b}=1}^n g^{\mathfrak{ab}}(\rho _{\odot}^\mathfrak{a},\rho _{\|}^\mathfrak{a},\rho ^\mathfrak{a}_{st}) \sqrt{\dot{\beta }^\mathfrak{a}\dot{\beta }^\mathfrak{b}},
\end{equation*}
where $g^{\mathfrak{ab}}(\rho _{\odot}^\mathfrak{a},\rho _{\|}^\mathfrak{a},\rho ^\mathfrak{a}_{st})$ is the $n\times n$ positive definite hardening matrix.

When the dissipation is taken into account, the above formulated variational principle must be modified. Following \citet{Sedov1965,Berdichevsky1967} we require that the true displacement field $\check{\mathbf{u}}(\mathbf{x})$ and the true plastic slips $\check{\beta }^\mathfrak{a}(\mathbf{x})$ in the {\it final} state of deformation in equilibrium obey the variational equation
\begin{equation}
\label{eq:2.8}
\delta I+\int_{\mathcal{V}}\sum_{\mathfrak{a}=1}^n \frac{\partial D}{\partial \dot{\beta }^\mathfrak{a}}\delta \beta ^\mathfrak{a}\, dx=0
\end{equation}
for all variations of admissible fields $\mathbf{u}(\mathbf{x})$ and $\beta ^\mathfrak{a}(\mathbf{x})$ satisfying the constraints \eqref{eq:2.2}. Together with the above formula for $\delta I$ and the arbitrariness of $\delta u_{i}$ and $\delta \beta ^\mathfrak{a}$ in $\mathcal{V}$ and on $\partial _s$, equation \eqref{eq:2.8} yields
\begin{equation}
\label{eq:2.9}
\sigma _{ij,j}=0, \quad s_i^\mathfrak{a}\kappa ^\mathfrak{a} _{ij,k}\epsilon _{ljk}m_l^\mathfrak{a}+\tau ^\mathfrak{a}=\frac{\partial D}{\partial \dot{\beta }^\mathfrak{a}}, \quad \mathfrak{a}=1,\ldots ,n
\end{equation}
which are subjected to the boundary conditions \eqref{eq:2.2} and \eqref{eq:2.6}. For the dissipation potential from \eqref{eq:2.7a} equations \eqref{eq:2.9}$_2$ become
\begin{equation}
\label{eq:2.10}
 s_i^\mathfrak{a}\kappa ^\mathfrak{a} _{ij,k}\epsilon _{ljk}m_l^\mathfrak{a}+\tau ^\mathfrak{a}=g^\mathfrak{a}(\rho _{\odot}^\mathfrak{a},\rho _{\|}^\mathfrak{a},\rho ^\mathfrak{a}_{st})\, \text{sign} \dot{\beta }^\mathfrak{a}, \quad \mathfrak{a}=1,\ldots ,n.
\end{equation}
According to \eqref{eq:2.10} the plastic slip $\beta ^\mathfrak{a}$ may evolve only
if the yield condition
\begin{equation}\label{eq:2.11}
| s_i^\mathfrak{a}\kappa ^\mathfrak{a} _{ij,k}\epsilon _{ljk}m_l^\mathfrak{a}+\tau ^\mathfrak{a}|=g^\mathfrak{a}(\rho _{\odot}^\mathfrak{a},\rho _{\|}^\mathfrak{a},\rho ^\mathfrak{a}_{st})
\end{equation}
is fulfilled. If $|s_i^\mathfrak{a}\kappa ^\mathfrak{a} _{ij,k}\epsilon _{ljk}m_l^\mathfrak{a}+\tau ^\mathfrak{a} |<g^\mathfrak{a}(\rho _{\odot}^\mathfrak{a},\rho _{\|}^\mathfrak{a},\rho ^\mathfrak{a}_{st})$ then $\beta ^\mathfrak{a}$ is ``frozen'':
$\dot{\beta }^\mathfrak{a}=0$. 

In the following we shall consider single crystals deforming in single slip and take the simplest expressions for the free energy density \citep{Berdichevsky2006a,Berdichevsky2006b}
\begin{equation}
\label{eq:2.12}
\psi =\frac{1}{2}\lambda (\text{tr} \bvarepsilon^e)^2+\mu \text{tr}(\bvarepsilon^e \cdot \bvarepsilon^e)+\mu k \ln \frac{1}{1-\rho /\rho _s},
\end{equation}
and dissipation
\begin{equation}
\label{eq:2.13}
D=(K+\mu \alpha b\sqrt{\rho +\rho _{st}})|\dot{\beta }|.
\end{equation}
In these formulas $\mu$ and $\lambda$ are Lam\'{e} elastic constants, $\rho_{s}$ the saturated dislocation density, $k$ and $\alpha $ are positive constants. It is assumed that only one sort of dislocation appears such that either $\rho _{\odot}=\rho $ (or $\rho _{\|}=\rho $). The second summand on the right-hand side of \eqref{eq:2.13} describes the Taylor hardening due to the GNDs and SSDs.

\section{Anti-plane constrained shear}
\label{sec:3}

\begin{figure}[htb]
	\centering
	\includegraphics[height=7cm]{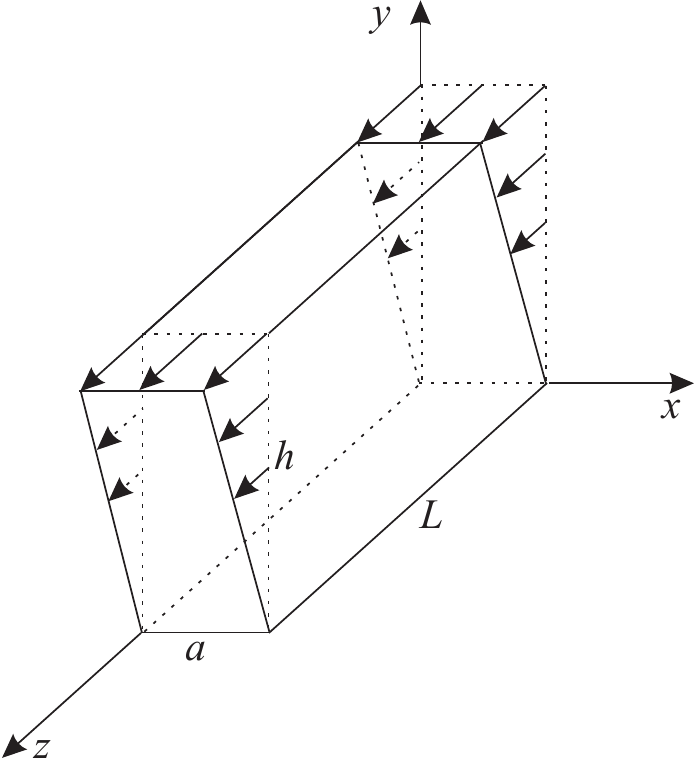}
	\caption{Anti-plane constrained
shear}
	\label{fig:Antiplaneshear}
\end{figure}

As an application of the proposed theory let us consider the single crystal layer undergoing an anti-plane shear deformation. Let $C$ be the cross section of the layer by planes $z=\text{const}$. For simplicity, we take $C$ as a rectangle of width $a$ and height $h$, $0<x\le a$, $0<y\le h$. We place this single crystal in a ``hard'' device with the prescribed displacement at the boundary $\partial C\times [0,L]$ (see Fig.~\ref{fig:Antiplaneshear})
\begin{equation*}
w =\gamma y \quad \text{at $\partial C\times [0,L]$},
\end{equation*}
where $w(x,y,z)$ is the $z$-component of the displacement and $\gamma $ corresponds to the overall shear strain. We may imagine this single crystal as a grain, where the hard device models the grain boundary. The height of the cross section, $h$, and the length of the beam, $L$, are assumed to be much larger than the width $a$ ($a\ll h$, $a\ll L$) to neglect the end effects and to have the stresses and strains depending only on one variable $x$ in the central part of the beam. If the shear strain is sufficiently small and the crystal is initially free of GNDs, then the crystal deforms elastically and $w=\gamma y$ everywhere in the specimen.
If $\gamma $ exceeds some critical value, then the screw GNDs may appear. We admit only one active slip system, with the slip planes parallel to the plane $y=0$ and the dislocation lines parallel to the $z$-axis. We aim at determining the distribution of GNDs as function of $\gamma $ within the framework of continuum dislocation theory taking into account the SSDs and Taylor hardening. For screw dislocations with the slip planes parallel to the plane $y=0$, the tensor of plastic distortion, $\beta _{ij}$, has only one non-zero component $\beta _{zy}\equiv \beta $. We assume that $\beta $ depends only on $x$-coordinate: $\beta =\beta (x)$. Since the displacements are prescribed at the boundary of the crystal, dislocations cannot penetrate the boundaries $x=0$ and $x=a$, therefore
\begin{equation}\label{eq:3.1}
\beta (0)=\beta (a)=0.
\end{equation}
The plastic strains are given by
\begin{equation*}
\varepsilon ^{(p)}_{yz}=\varepsilon ^{(p)}_{zy}=\frac{1}{2}\beta
(x).
\end{equation*}
The only non-zero component of the tensor of dislocation density,
$\alpha _{ij}=\varepsilon _{jkl}\beta _{il,k}$, is
\begin{equation*}
\alpha _{zz}=\beta _{,x},
\end{equation*}
so the number of GNDs per unit area perpendicular to the $z$-axis is $\rho =|\beta _{,x}|/b$. Under the assumptions made, the energy density per unit volume of the crystal with dislocations \eqref{eq:2.12} takes a simple form
\begin{equation}\label{eq:3.3}
U=\frac{1}{2}\mu (\gamma -\beta )^2+\mu k \ln \frac{1}{1-|\beta
_{,x}|/\rho _s b},
\end{equation}
while the dissipation potential becomes
\begin{equation}
\label{eq:3.4}
D=\left( K+\mu \alpha b\sqrt{|\beta _{,x}|/b+\rho _{st}}\right) |\dot{\beta }|.
\end{equation}

We first analyze the dislocation nucleation and dislocation pile-up during the loading when $\gamma $ increases from zero so that $\dot{\beta }\ge 0$. In this case $\beta (x)$ must satisfy the variational equation
\begin{equation}
\begin{split}
\label{eq:3.5}
\delta \int_{0}^a \left[ \frac{1}{2}\mu (\gamma -\beta )^2+\mu k \ln \frac{1}{1-|\beta
_{,x}|/\rho _s b}\right] \, dx
\\
+\int_{0}^a \left( K+\mu \alpha b\sqrt{|\beta _{,x}|/b+\rho _{st}}\right)  \delta \beta \, dx=0.
\end{split}
\end{equation}
It is convenient to divide equation \eqref{eq:3.5} by $\mu $ and introduce the following dimensionless quantities
\begin{equation}\label{eq:3.6}
\bar{x} =xb\rho _s, \quad \bar{a}=ab\rho _s,\quad \gamma _0=\frac{K}{\mu },\quad \xi =\alpha b\sqrt{\rho _s}, \quad \kappa =\frac{\rho _{st}}{\rho _s}.
\end{equation}
Since we shall deal with these dimensionless quantities only, the bar over them can be dropped for short. Equation \eqref{eq:3.5} reduces to
\begin{equation}\label{eq:3.7}
\delta \int_{0}^a \left[ \frac{1}{2}(\gamma -\beta )^2+k \ln \frac{1}{1-|\beta ^\prime |}\right] \, dx
+\int_{0}^a \left( \gamma _0+\xi \sqrt{|\beta ^\prime |+\kappa }\right)  \delta \beta \, dx=0,
\end{equation}
with prime denoting the derivative.

Due to the boundary conditions \eqref{eq:3.1} $\beta ^\prime (x)$ should change its sign
on the interval $(0,a)$. One-dimensional theory of dislocation pile-ups \citep{Berdichevsky2007,Kochmann2008,Kochmann2009a,Le2008a,Le2008b,Le2009} suggests to seek the solution in the form
\begin{equation}\label{eq:3.8}
\beta (x)=
  \begin{cases}
    \beta _1(x) & \text{for $x \in (0,l)$}, \\
    \beta _m & \text{for $x \in (l,a-l)$}, \\
    \beta _1(a-x ) & \text{for $x \in (a-l,
    a)$},
  \end{cases}
\end{equation}
where $\beta _m$ is a constant, $l$ an unknown length, $0\le l\le
a/2$, and $\beta _1(l)=\beta _m$ at $x=l$. We have to find $\beta _1(x)$
and the constants, $\beta _m$ and $l$. Since $\beta _1^\prime >0$ for $x \in
(0,l)$, the first integral in \eqref{eq:3.7} with this Ansatz reads
\begin{equation*}
2\int_0^l\left[ \frac{1}{2}(\gamma -\beta _1)^2+k \ln \frac{1}{1-\beta _1^\prime }\right] dx +\frac{1}{2}(\gamma -\beta _m)^2(a -2l).
\end{equation*}
Likewise the last integral in \eqref{eq:3.7} becomes
\begin{equation*}
2\int_0^l \left( \gamma _0+\xi \sqrt{\beta _1^\prime +\kappa }\right) \delta \beta _1 \, dx +(\gamma _0+\xi \sqrt{\kappa })(a-2l)\, \delta \beta _m.
\end{equation*}
Varying the first integral with respect to $\beta _1(x)$ and substituting the result into \eqref{eq:3.7} we obtain from it the equation for $\beta _1(x)$ on the interval $x \in (0,l)$
\begin{equation}\label{eq:3.9}
\gamma - \gamma _0-\xi \sqrt{\beta _1^\prime +\kappa }-\beta _1+\frac{k\beta _1^{\prime \prime }}{(1-\beta _1^\prime )^2}=0.
\end{equation}
The variation with respect to $\beta _m$ and $l$ yields the two additional boundary conditions at $x =l$
\begin{equation}\label{eq:3.10}
\beta _{1}^\prime (l)=0, \quad 2k-(\gamma _r -\beta _m)(a-2l)=0,
\end{equation}
with
\begin{equation}\label{eq:3.10a}
\gamma _r=\gamma -\gamma _0-\xi \sqrt{\kappa }.
\end{equation}
Condition \eqref{eq:3.10}$_1$ means that the dislocation density must be continuous.

At the onset of dislocation nucleation the dimensionless density of GNDs $\beta ^\prime $ must be small compared to 1, so equation \eqref{eq:3.9} can be replaced by
\begin{equation}\label{eq:3.11}
\gamma _r -\beta _1+k\beta _1^{\prime \prime }=0.
\end{equation}
Together with the boundary conditions \eqref{eq:3.1}$_1$ and \eqref{eq:3.10}$_1$ this yields
\begin{equation*}
\beta _1(x)=\gamma _r (1 -\cosh \frac{x}{\sqrt{k}}+\tanh
\frac{l}{\sqrt{k}}\sinh \frac{x}{\sqrt{k}}),\quad 0\le x \le
l.
\end{equation*}
Consequently, \eqref{eq:3.10}$_2$ gives the following transcendental equation to determine $l$ in terms of the constants $k$ and $a$
\begin{equation}\label{eq:3.12}
f(l)\equiv 2l+2\frac{k}{\gamma _r}\cosh \frac{l}{\sqrt{k}}=a.
\end{equation}
According to \eqref{eq:3.8} $l$ must lie in the segment $[0,a/2]$. Since $\cosh (l/\sqrt{k})\ge 1$, $2l\le a-2k/\gamma _r$. Thus, equation \eqref{eq:3.12} has no positive root if $\gamma _r\le 2k/a$. Returning to the original variables according to \eqref{eq:3.6} and \eqref{eq:3.10a} we see that inequality $\gamma _r \le 2k/a$ corresponds to the condition $\gamma \le \gamma _c$, where
$$\gamma _c=\gamma _0 +\xi \sqrt{\kappa }+\frac{2k}{a},$$
and for $\gamma \le \gamma _{c}$  no dislocations are nucleated. This formula resembles the combined \citet{Taylor1934} and \citet{Hall1951,Petch1953} relations. 

Assume now that at some stage of loading after the nucleation and accumulation of GNDs we stop increasing $\gamma $ and then decrease it. The plastic slip rate could be either zero or negative. If the yield condition \eqref{eq:2.11} cannot be fulfilled, then the plastic slip rate must be zero and the plastic slip $\beta (x)$ is frozen during this process. We call such process during which $\gamma $ decreases but $\beta (x)$ remains unchanged and equal to that plastic slip $\beta (x)$ at the end of the loading process elastic unloading. We will see that the average stress depends linearly on $\gamma $ during this elastic unloading. It remains now to analyze the last case $\dot{\beta }<0$ which we call loading in the opposite direction. In this case the variational equation \eqref{eq:3.7} changes to
\begin{equation}\label{eq:3.13}
\delta \int_{0}^a [\frac{1}{2}(\gamma -\beta )^2+k \ln \frac{1}{1-|\beta ^\prime |}]\, dx
-\int_{0}^a (\gamma _0+\xi \sqrt{|\beta ^\prime |+\kappa })\, \delta \beta \, dx=0,
\end{equation}
In the first stage of loading in the opposite direction the distribution of GNDs should not change much as compared to that at the end of the loading process. Therefore we can still assume $\beta (x)$ in the form \eqref{eq:3.8}. With this solution Ansatz it is straightforward to derive from \eqref{eq:3.13} the following equation
\begin{equation}\label{eq:3.14}
\gamma + \gamma _0+\xi \sqrt{\beta _1^\prime +\kappa }-\beta _1+\frac{k\beta _1^{\prime \prime }}{(1-\beta _1^\prime )^2}=0.
\end{equation}
and the boundary conditions
\begin{equation}\label{eq:3.15}
\beta _{1}^\prime (l)=0, \quad 2k=(\gamma _l -\beta _m)(a-2l),
\end{equation}
where
\begin{equation*}
\gamma _l=\gamma +\gamma _0+\xi \sqrt{\kappa }.
\end{equation*}
If the dimensionless density of GNDs $\beta ^\prime $ is small compared to 1 at the end of the loading in the opposite direction, equation \eqref{eq:3.14} can approximately be replaced by
\begin{equation}\label{eq:3.16}
\gamma _l -\beta _1+k\beta _1^{\prime \prime }=0.
\end{equation}
In this linearized version the system of equation \eqref{eq:3.16} and boundary conditions \eqref{eq:3.15} are identical with the system \eqref{eq:3.11} and \eqref{eq:3.10} if $\gamma _l$ is replaced by $\gamma _r$. Consequently, the solutions of these problems are equal if $\gamma _l=\gamma _r$. In particular, the GNDs are completely annihilated at 
$$\gamma = -\gamma _c=-\gamma _0 -\xi \sqrt{\kappa }-\frac{2k}{a}.$$

\section{Numerical simulations and comparison}
\label{sec:4}

\begin{figure}[htb]
	\centering
	\includegraphics[width=8cm]{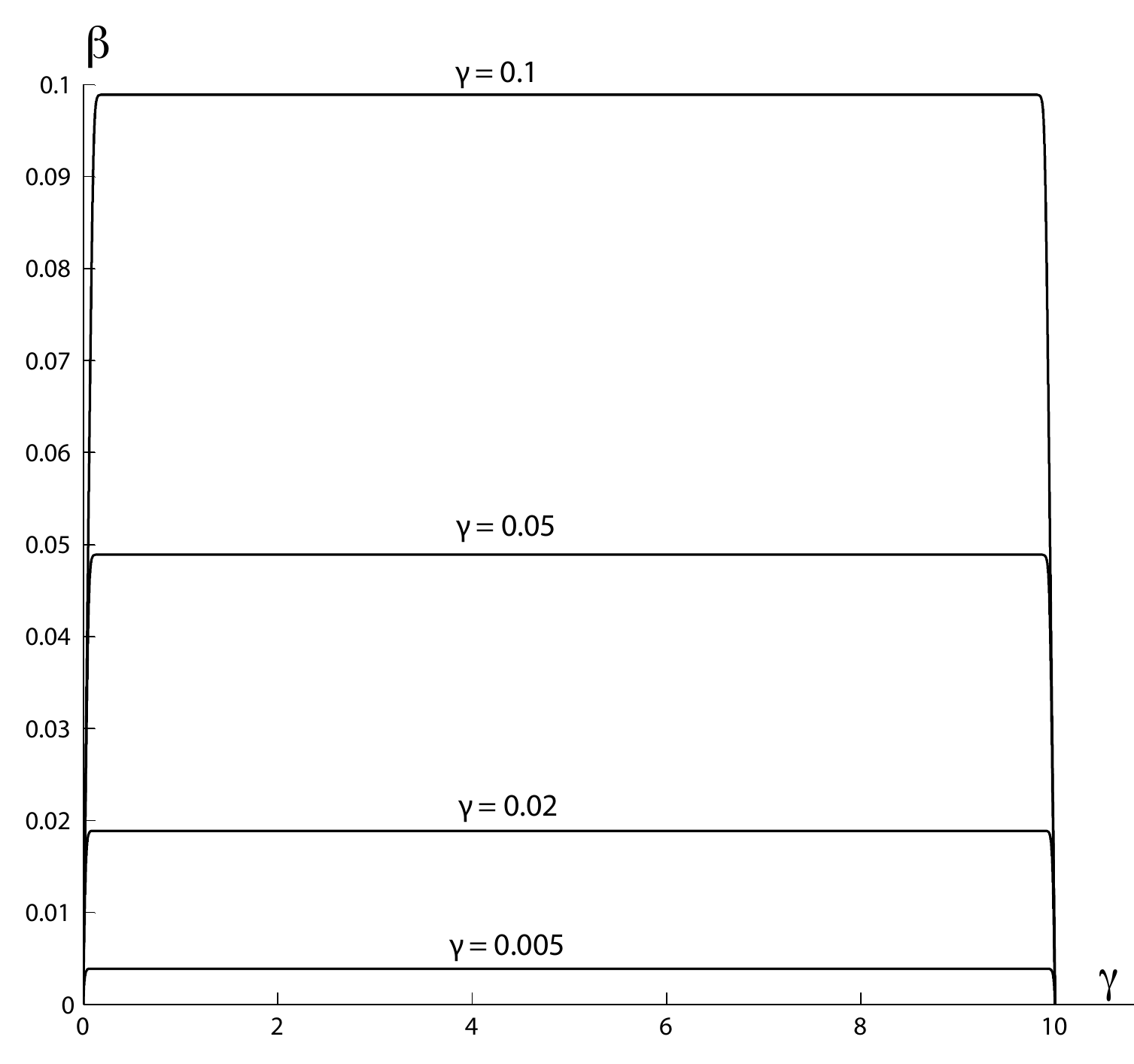}
	\caption{Evolution of $\beta $: 
a) $\gamma =0.005$, b) $\gamma =0.02$, c) $\gamma =0.05$, d) $\gamma =0.05$}
	\label{fig:beta}
\end{figure}

\begin{figure}[htb]
	\centering
	\includegraphics[width=8cm]{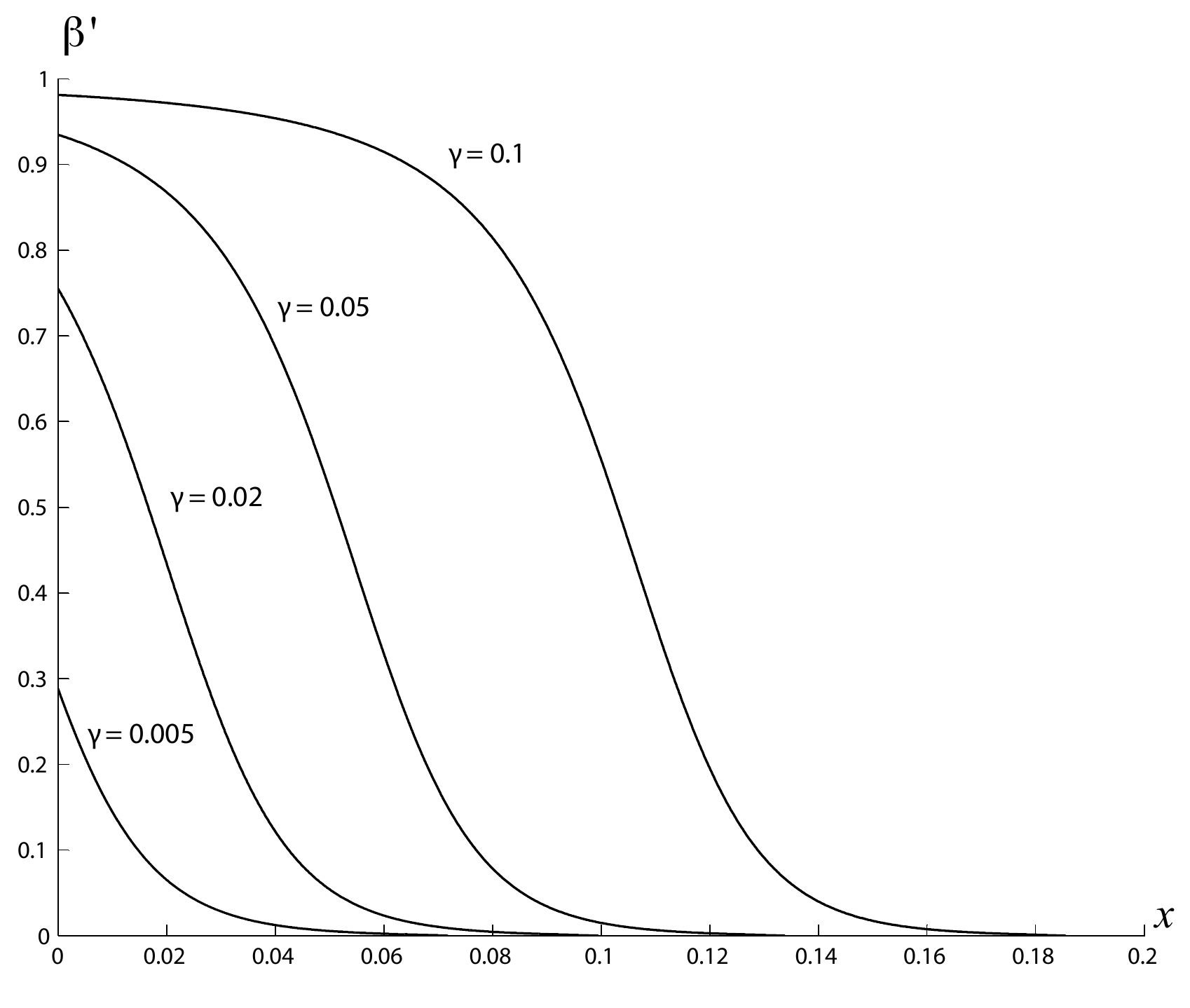}
	\caption{Evolution of $\beta ^\prime _1(x)$ in the left boundary layer: 
a) $\gamma =0.005$, b) $\gamma =0.02$, c) $\gamma =0.05$, d) $\gamma =0.05$}
	\label{fig:density}
\end{figure}

In general the two-point boundary-value problem \eqref{eq:3.9}, \eqref{eq:3.1}$_1$, and \eqref{eq:3.10}$_1$ for any given positive $l\in (0,a/2)$, due to its nonlinearity, can only be solved numerically. Since the slope of $\beta (x)$ at $x=0$ is {\it a priori} unknown, we use the collocation method and employ the Matlab program bvp4c for the solution of this two-point boundary-value problem \citep{Kirzenka2001,Shampine2003}. The third term of the equation \eqref{eq:3.9} could lead sometimes to the complex solution if $\beta _1^\prime +\kappa $ suggested by the collocation method during the iteration turns out to be negative. Therefore we modify it to be $-\xi \sqrt{|\beta _1^\prime |+\kappa }$. For $\gamma $ close to the threshold value where $\beta ^\prime (x)$ is small, we use the solution of the linearized problem found above as the initial guess for the solution of the nonlinear equation. For larger $\gamma $ we solve the problem incrementally and use the solution obtained at the previous step as the initial guess for the solution at the next $\gamma $ differing from the previous one by a small increment. With this numerical solution at hand we can determine the function $\beta _m(l)=\beta _1(l)$ so that equation \eqref{eq:3.10}$_2$ can be solved numerically. The latter is solved by the bisection method which is robust because it does not require the derivative of function on the left-hand side of \eqref{eq:3.10}$_2$. Fig.~\ref{fig:beta} shows the evolution of the plastic slip $\beta (x)$ at different values of the overall shear $\gamma $. For the numerical simulation we took $k=10^{-4}$, $\rho _s=10^{15}$m$^{-2}$, $b=10^{-10}$m, $a=10^{-4}$m, so that $\bar{a}=ab\rho _s=10$. Other material parameters are chosen such that $\xi =10^{-3}$, $\kappa =10^{-2}$, and $\gamma _0=K/\mu =10^{-3}$. From this Figure one can see that the plastic slip $\beta (x)$ increases as $\gamma $ increases. There are two boundary layers in which the geometrically necessary dislocations pile up against the obstacles at $x=0$ and $x=a$. In the middle $(l,a-l)$ the crystal is free of GNDs.

\begin{figure}[htb]
	\centering
	\includegraphics[width=8cm]{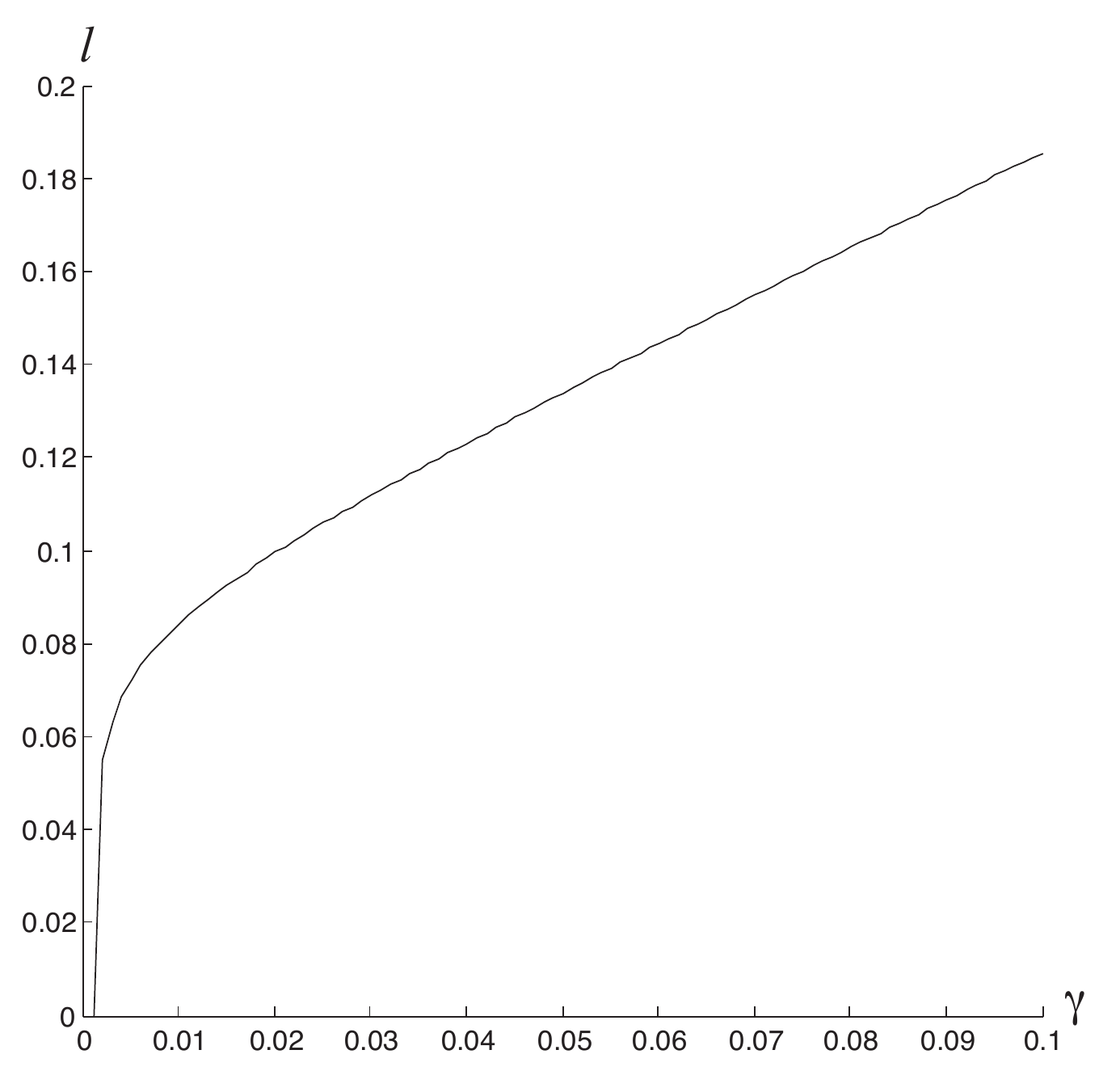}
	\caption{Thickness of the boundary layer $l$ as function of $\gamma $}
	\label{fig:length}
\end{figure}

Fig.~\ref{fig:density} shows the evolution of the dimensionless density of geometrically necessary dislocations (GNDs) $\beta ^\prime _1(x)$ in the left boundary layer as $\gamma $ increases. We see that the density of GNDs is a monotonously decreasing function with the maximum being achieved at $x=0$. As $\gamma $ increases, the number of GNDs also increase and they pile up against the obstacle at $x=0$. At $\gamma =0.1$ the density of GNDs is already close to the saturated dislocation density. The dependence of the thickness of boundary layer $l$ (where the GNDs are distributed) on the overall shear $\gamma $ is shown in Fig.~\ref{fig:length}. The thickness $l$ starts from zero at $\gamma =\gamma _c$ and increases monotonically as $\gamma $ increases. For moderate and large values of $\gamma $ the thickness $l$ is nearly a linear function of $\gamma $.

\begin{figure}[htb]
	\centering
	\includegraphics[width=8cm]{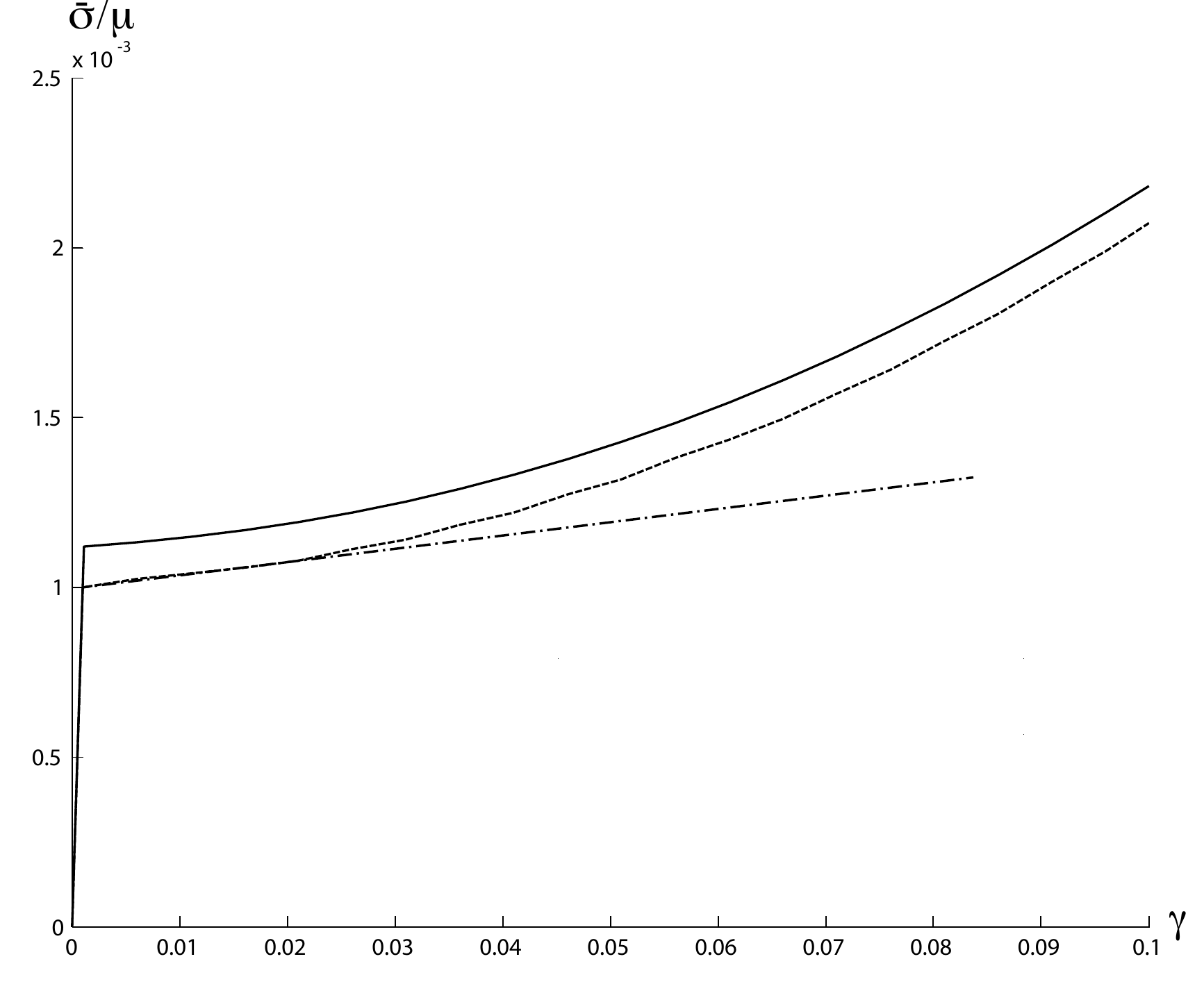}
	\caption{Normalized average stress versus shear strain curve: i) theory with SSDs and Taylor hardening (bold line), ii) theory without SSDs and Taylor hardening (dashed line), iii) linearized theory without SSDs and Taylor hardening (dashed and dotted line)}
	\label{fig:stress-strain}
\end{figure}

As soon as the plastic slip develops, the shear stress $\sigma =\mu (\gamma -\beta (x))$ becomes inhomogeneous. It is interesting to plot the average shear stress
\begin{equation}\label{eq:stress}
\bar{\sigma }=\frac{1}{a}\int_0^a\mu (\gamma -\beta (x))\, dx
\end{equation}
as function of the shear strain. For $\gamma \le \gamma _{c}$ the plastic slip $\beta (x)=0$, so $\bar{\sigma }=\mu \gamma $. For $\gamma > \gamma _{c}$ the average stress will be less than $\mu \gamma $ due to the positiveness of the plastic slip. However, one can still observe the positive slope of the stress-strain curve which can be explained by the kinematic and Taylor hardening in combination. Fig.~\ref{fig:stress-strain} shows the normalized average shear stress $\bar{\sigma }/\mu $ versus the shear strain curve (bold line). As comparison, we show also the stress-strain curve obtained in the continuum dislocation theory without SSDs and Taylor hardening (dashed line) as well as the linearized theory without SSDs and Taylor hardening (dashed and dotted line). One can see that the slope of the stress-strain curve obtained by the CDT taking into account SSDs and Taylor hardening is highest. Mention that the threshold stress at which the GNDs begin to nucleate in the theory without SSDs and Taylor hardening is $\gamma _0+2k/a$ that is also lower than the threshold value $\gamma _c$.

\begin{figure}[htb]
	\centering
	\includegraphics[width=8cm]{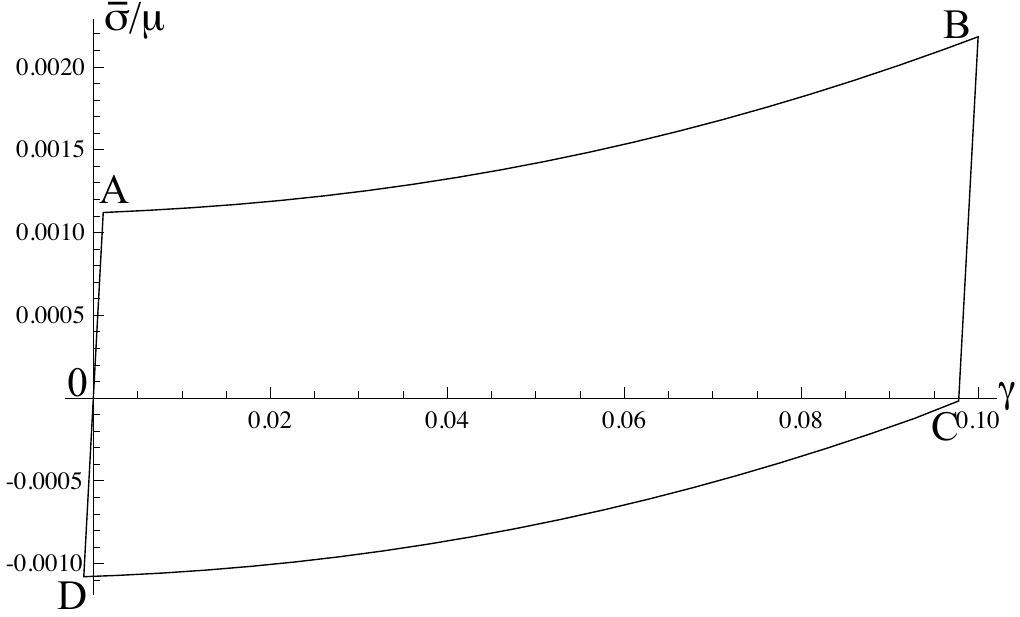}
	\caption{The stress-strain curve during loading, elastic loading and unloading, and loading in the opposite direction}
	\label{fig:combined}
\end{figure}

On Fig.~\ref{fig:combined} we show the normalized stress-strain curve obtained during loading, elastic loading and unloading, and loading in the opposite direction. The curve AB corresponds to the stress-strain curve during the loading process when $\gamma $ increases from $\gamma _c$. The straight lines DA going through the origin and BC are the stress-strain curves during elastic loading and unloading when $\gamma $ increases (decreases) at frozen $\beta $. It can be seen from \eqref{eq:stress} that the slope of DA and BC must be constant because of the frozen $\beta (x)$. Finally, the curve CD corresponds to the stress-strain curve during the loading in the opposite direction when the plastic slip $\beta (x)$ decreases simultaneously with the decrease of $\gamma $. This curve is obtained from \eqref{eq:stress} where the plastic slip $\beta (x)$ corresponds to the solution of the problem \eqref{eq:3.14} and \eqref{eq:3.15}. The latter is solved numerically with the use of Matlab bvp4c in the similar manner as for the system \eqref{eq:3.9} and \eqref{eq:3.10}. This stress strain curve cuts the elastic line AD at point D with the coordinate $(-\gamma _c,-\gamma _c)$ at which the plastic slip $\beta (x)$ must be equal to zero.

\begin{figure}[htb]
	\centering
	\includegraphics[width=7cm]{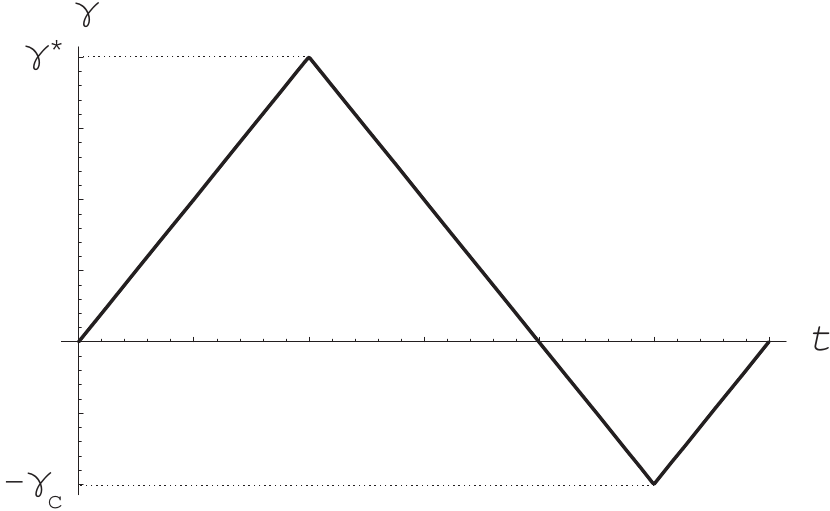}
	\caption{Loading path}
	\label{fig:LoadingPath}
\end{figure}

Let us consider now the following close loading path: $\gamma $ is first increased from zero to some value $\gamma _*>\gamma _c$, then decreased to $-\gamma _c=-\gamma _0 -\xi \sqrt{\kappa }-\frac{2k}{a}$, and finally increased to zero (Fig.~\ref{fig:LoadingPath}). The rate of change of $\gamma (t)$ does not affect the results due to the rate independence of the dissipation. 

In Fig.~\ref{fig:combined} the close loop OABCDO shows the normalized average shear stress (or average elastic shear strain) versus shear strain curve for the loading program of Fig.~\ref{fig:LoadingPath}. The straight line OA corresponds to the purely elastic loading with $\gamma $ increasing from zero to $\gamma _c$, where $\beta (x)=0$. The line AB corresponds to the plastic yielding where the yield condition \eqref{eq:2.11} (or, equivalently, the system \eqref{eq:3.9} and \eqref{eq:3.10}) holds true. The yield begins at the
point A with the yield stress $\sigma _y=\mu \gamma _c$. The work hardening
due to the nucleation and pile-up of GNDs and Taylor hardening is observed. During the unloading as $\gamma $ decreases from $\gamma ^*$ to $\gamma _*=\gamma ^*-2\gamma _c $ (the line BC) the plastic slip $\beta =\beta ^*$ is frozen. As $\gamma $ decreases further from $\gamma _*$ to $-\gamma _c$, the plastic yielding occurs where the yield condition \eqref{eq:2.11} (or, equivalently, the system \eqref{eq:3.14} and \eqref{eq:3.15}) holds true (the line CD). The yield stress $\sigma _y=\mu (\gamma ^*-2\gamma _c)$ at the point C, at which the inverse plastic flow sets on, is larger than $-\mu \gamma _c$ (because $\gamma ^*>\gamma _c $). Along the line CD, as $\gamma $ is decreased, the created GNDs annihilate,
and at the point D all GNDs disappear. Finally, as $\gamma $ increases from $-\gamma _c$ to zero, the crystal behaves elastically with $\beta =0$. In this close cycle ABCD dissipation
occurs only on the lines AB and CD. It is interesting that the lines DA and BC are parallel and have the same length. In phenomenological plasticity theory this property is modeled as
the translational shift of the yield surface in the stress space,
the so-called Bauschinger effect.

\section{Conclusion}
\label{sec:5}
In this paper we have proposed the continuum dislocation theory taking into account the statistically stored dislocations and Taylor hardening. On the example of the anti-plane constrained shear we have shown that the threshold stress at which the geometrically necessary dislocations nucleate is higher than that of the theory without SSDs and Taylor hardening. For the loading path consisting of increasing and decreasing shear strain
the stress-strain curve becomes a hysteresis loop. The work hardening and the Bauschinger effect are quantitatively described in terms of the plastic slip and the densities of GNDs and SSDs.

\end{document}

%% file: def.tex
\newcommand{\balpha}{\boldsymbol{\alpha}}
\newcommand{\bbeta}{\boldsymbol{\beta}}
\newcommand{\bgamma}{\boldsymbol{\gamma}}
\newcommand{\bdelta}{\boldsymbol{\delta}}
\newcommand{\bepsilon}{\boldsymbol{\epsilon}}
\newcommand{\bvarepsilon}{\boldsymbol{\varepsilon}}
\newcommand{\bzeta}{\boldsymbol{\zeta}}
\newcommand{\boldeta}{\boldsymbol{\eta}}
\newcommand{\btheta}{\boldsymbol{\theta}}
\newcommand{\bvartheta}{\boldsymbol{\vartheta}}
\newcommand{\biota}{\boldsymbol{\iota}}
\newcommand{\bkappa}{\boldsymbol{\kappa}}
\newcommand{\blambda}{\boldsymbol{\lambda}}
\newcommand{\bmu}{\boldsymbol{\mu}}
\newcommand{\bnu}{\boldsymbol{\nu}}
\newcommand{\bxi}{\boldsymbol{\xi}}
\newcommand{\bpi}{\boldsymbol{\pi}}
\newcommand{\bvarpi}{\boldsymbol{\varpi}}
\newcommand{\brho}{\boldsymbol{\rho}}
\newcommand{\bvarrho}{\boldsymbol{\varrho}}
\newcommand{\bsigma}{\boldsymbol{\sigma}}
\newcommand{\bvarsigma}{\boldsymbol{\varsigma}}
\newcommand{\btau}{\boldsymbol{\tau}}
\newcommand{\bupsilon}{\boldsymbol{\upsilon}}
\newcommand{\bphi}{\boldsymbol{\phi}}
\newcommand{\bvarphi}{\boldsymbol{\varphi}}
\newcommand{\bchi}{\boldsymbol{\chi}}
\newcommand{\bpsi}{\boldsymbol{\psi}}
\newcommand{\bomega}{\boldsymbol{\omega}}
\newcommand{\bGamma}{\boldsymbol{\Gamma}}
\newcommand{\bDelta}{\boldsymbol{\Delta}}
\newcommand{\bTheta}{\boldsymbol{\Theta}}
\newcommand{\bLambda}{\boldsymbol{\Lambda}}
\newcommand{\bXi}{\boldsymbol{\Xi}}
\newcommand{\bPi}{\boldsymbol{\Pi}}
\newcommand{\bSigma}{\boldsymbol{\Sigma}}
\newcommand{\bUpsilon}{\boldsymbol{\Upsilon}}
\newcommand{\bPhi}{\boldsymbol{\Phi}}
\newcommand{\bPsi}{\boldsymbol{\Psi}}
\newcommand{\bOmega}{\boldsymbol{\Omega}}
\newcommand{\llbracket}{[\![}
\newcommand{\rrbracket}{]\!]}

%% file: SsdTaylor.bbl
\begin{thebibliography}{10}

\bibitem[Acharya and Bassani(2001)]{Acharya2001}
Acharya, A., Bassani, J.L., 2001. Lattice incompatibility and a gradient theory of
crystal plasticity. Journal of the Mechanics and Physics of Solids 48, 1565-1595.
\bibitem[Arsenlis and Parks(1999)]{Arsenlis1999}
Arsenlis, A., Parks, D.M., 1999. Crystallographic aspects of geometrically-necessary and statistically-stored dislocation density. Acta Materialia 47(5), 1597-1611.
\bibitem[Arsenlis et~al.(2004)]{Arsenlis2004}
Arsenlis, A., Parks, D.M., Becker, R., Bulatov, V.V., 2004. On the evolution of crystallographic dislocation density in non-homogeneously deforming crystals. Journal of the Mechanics and Physics of Solids, 52(6), 1213-1246.
\bibitem[Ashby(1970)]{Ashby1970} Ashby, M.F., 1970. The deformation of plastically non-homogeneous materials. Philosophical Magazine 21, 399-424.
\bibitem[Berdichevsky and Sedov(1967)]{Berdichevsky1967} Berdichevskii, V.L., Sedov, L.I., 1967. Dynamic theory of continuously distributed dislocations. Its relation to plasticity theory. Journal of Applied Mathematics and Mechanics 31(6), 989-1006.
\bibitem[Berdichevsky(2006a)]{Berdichevsky2006a}
Berdichevsky, V.L.,  2006a. Continuum theory of dislocations revisited.
Continuum Mech. Thermodyn. 18, 195-222.
\bibitem[Berdichevsky(2006b)]{Berdichevsky2006b}
Berdichevsky, V.L.,  2006b. On thermodynamics of crystal plasticity.
 Scripta Materialia  54, 711-716.
\bibitem[Berdichevsky and Le(2002)]{Berdichevsky2002}
Berdichevsky, V.L., Le, K.C., 2002. Theory of charge nucleation in two dimensions. Physical Review E 66(2), 026129.
\bibitem[Berdichevsky and Le(2007)]{Berdichevsky2007}
Berdichevsky, V.L.,  Le, K.C., 2007. Dislocation nucleation and work hardening in
  anti-planed constrained shear. Continuum Mech. Thermodyn.  18, 455-467.
\bibitem[Bilby(1955)]{Bilby1955}
Bilby, B., 1955. Types of dislocation source, in: Report of Bristol Conference on Defects in Crystalline Solids (Bristol 1954, London: The Physical Soc.), pp. 124-133.
\bibitem[Gurtin(2002)]{Gurtin2002}
Gurtin, M.E., 2002. A gradient theory of single-crystal viscoplasticity that accounts for geometrically necessary dislocations. Journal of the Mechanics and Physics of Solids 50, 5-32.
\bibitem[Gurtin et~al.(2007)]{Gurtin2007}
Gurtin, M.E., Anand, L., Lele, S.P., 2007. Gradient single-crystal plasticity with free energy dependent on dislocation densities. Journal of the Mechanics and Physics of Solids 55(9), 1853-1878.
\bibitem[Hall(1951)]{Hall1951} Hall, E.O., 1951. The deformation and ageing of mild steel. Proc. Phys. Soc. B 64, 742-753. 
\bibitem[Hansen and Kuhlmann-Wilsdorf(1986)]{Hansen1986}
Hansen, N., Kuhlmann-Wilsdorf, D., 1986. Low energy dislocation structures due to unidirectional deformation at low temperatures. Materials Science and Engineering 81, 141-161.
\bibitem[Hirth and Lothe (1968)]{Hirth1968}
Hirth, J.P., Lothe, J., 1968. Theory of dislocations. McGraw-Hill, New York.
\bibitem[Kaluza and Le(2011)]{Kaluza2011}
Kaluza, M.,  Le, K.C.,  2011. On torsion of a single crystal rod. International Journal of Plasticity  27, 460-469.
\bibitem[Kirzenka(2001)]{Kirzenka2001}
Kierzenka, J., Shampine, L.F., 2001. A BVP solver based on residual control and the Matlab PSE. ACM Transactions on Mathematical Software (TOMS) 27(3): 299-316.
\bibitem[Kochmann and Le(2008)]{Kochmann2008}
Kochmann, D.M.,  Le, K.C., 2008. Dislocaton pile-ups in bicrystals within continuum
  dislocation theory. International Journal of Plasticity  24, 2125-2147.
\bibitem[Kochmann and Le(2009a)]{Kochmann2009a}
Kochmann, D.M.,  Le, K.C., 2009. Plastic deformation of bicrystals within continuum
dislocation theory. Mathematic and Mechanics of Solids 14, 540-563.
\bibitem[Kochmann and Le(2009b)]{Kochmann2009b}
Kochmann, D.M.,  Le, K.C., 2009. A continuum model for initiation and evolution of
deformation twinning. Journal of the Mechanics and Physics of Solids  57, 987-1002.
\bibitem[Koster et al.(2015)]{Koster2015}  
Koster, M., Le, K.C., Nguyen, B.D., 2015. Formation of grain boundaries in ductile single crystals at finite plastic deformations. International Journal of Plasticity, DOI: 10.1016/j.ijplas.2015.02.010.
\bibitem[Kr{\"o}ner(1955)]{Kroener1955}
Kr{\"o}ner, E., 1955.
 {Der fundamentale Zusammenhang zwischen
  Versetzungsdichte und Spannungsfunktionen}.
 Zeitschrift f{\"u}r Physik A Hadrons and Nuclei
   142,  463--475.
\bibitem[Kr{\"o}ner(1992)]{Kroener1992}
 Kr{\"o}ner, E.,  1992. Mikrostrukturmechanik. GAMM-Mitteilungen 15, 104-119.
\bibitem[Kuhlmann-Wilsdorf(1989)]{Kuhlmann1989}
Kuhlmann-Wilsdorf, D. 1989. Theory of plastic deformation:-properties of low energy dislocation structures. Materials Science and Engineering A113, 1-41.
\bibitem[Kysar et al.(2010)]{Kysar2010}
Kysar, J.W., Saito, Y., \"Oztop, M.S., Lee, D., Huh, W.T., 2010. 
Experimental lower bounds on geometrically necessary dislocation density. 
International Journal of Plasticity 26, 1097-1123.
\bibitem[Laird et~al.(1986)]{Laird1986}
Laird, C., Charsley, P., Mughrabi, H., 1986. Low energy dislocation structures produced by cyclic deformation. Materials Science and Engineering 81, 433-450.
\bibitem[Le and G{\"u}nther(2014)]{Le2014}  
Le, K.C., G\"unther, C., 2014. Nonlinear continuum dislocation theory revisited. International Journal of Plasticity, 53, 164-178.
\bibitem[Le and Nguyen(2012)]{Le2012}
Le, K.C.,  Nguyen, B.D., 2012. Polygonization: Theory and comparison with
experiments. International Journal of Engineering Science  59, 211-218.
\bibitem[Le and Nguyen(2013)]{Le2013}
Le, K.C., Nguyen, B.D., 2013. 
 On bending of single crystal beam with continuously distributed dislocations. 
International Journal of Plasticity 48, 152-167.
\bibitem[Le and Nguyen(2010a)]{Le2010a}
Le, K.C.,  Nguyen, Q.S., 2010. Polygonization as low energy dislocation structure.
 Continuum Mechanics and Thermodynamics  22, 291-298.
\bibitem[Le and Sembiring(2008a)]{Le2008a}
Le, K.C.,  Sembiring, P., 2008a. Plane-constrained shear of a single crystal strip
  with two active slip-systems. Journal of the Mechanics and Physics of Solids  56,
   2541-2554.
\bibitem[Le and Sembiring(2008b)]{Le2008b}
Le, K.C.,  Sembiring, P., 2008b. Plane constrained shear of single crystals within
  continuum dislocation theory. Archive of Applied Mechanics 78, 587-597.
\bibitem[Le and Sembiring(2009)]{Le2009}
Le, K.C.,  Sembiring, P., 2009. Plane constrained uniaxial extension of a single
  crystal strip. International Journal of Plasticity 25, 1950-1969.
\bibitem[Le and Stumpf(1996a)]{Le1996a}
Le, K.C.,  Stumpf, H., 1996a. A model of elastoplastic bodies with continuously
distributed dislocations. International Journal of Plasticity 12, 611-627.
\bibitem[Le and Stumpf(1996b)]{Le1996b}
Le, K.C.,  Stumpf, H., 1996b. Nonlinear continuum theory of dislocations.
International Journal of Engineering Science  34, 339-358.
\bibitem[Le and Stumpf(1996c)]{Le1996c}
Le, K.C.,  Stumpf, H., 1996c. On the determination of the crystal reference in
nonlinear continuum theory of dislocations. Proceedings of the Royal Society of London  A452, 359-371.
\bibitem[Nye(1953)]{Nye1953}
Nye, J.F., 1953. Some geometrical relations in dislocated crystals. Acta metallurgica 1(2), 153-162.
\bibitem[Ortiz and Repetto(1999)]{Ortiz1999}
Ortiz, M.,  Repetto, E.A., 1999. Nonconvex energy minimization and dislocation
structures in ductile single crystals. Journal of the Mechanics and Physics of Solids  47,
397-462.
\bibitem[Ortiz et~al.(2000)]{Ortiz2000}
Ortiz, M.,  Repetto, E.A., Stainier, L.,  2000. A theory of subgrain dislocation structures.
Journal of the Mechanics and Physics of Solids  48, 2077-2114.
\bibitem[Petch(1953)]{Petch1953} 
Petch, N.J., 1953. The cleavage strength of polycrystals, Journal of the Iron and Steel Institute 174, 25--28.
\bibitem[Puglisi and Truskinovsky(2005)]{Puglisi2005}
Puglisi, G., Truskinovsky, L., 2005. Thermodynamics of rate-independent plasticity. Journal of the Mechanics and Physics of Solids 53(3), 655-679.
\bibitem[Taylor(1934)]{Taylor1934} 
Taylor, G.I., 1934. The mechanism of plastic deformation of crystals. Proc. Roy. Soc. London A 145, 362-387.
\bibitem[Sedov(1965)]{Sedov1965} 
Sedov, L.I., 1965. Mathematical methods for constructing new models of continuous media. Russian Mathematical Surveys, 20(5): 123--182 (1965).
\bibitem[Shampine et~al.(2003)]{Shampine2003}
Shampine, L.F., Gladwell, I., Thompson, S., 2003. Solving ODEs with Matlab. Cambridge University Press, Cambridge. 

\end{thebibliography}
